\documentclass[aps,prl,twocolumn,superscriptaddress]{revtex4}
\usepackage{graphicx,epsf}
\usepackage{amsmath}
\usepackage{color}

\newcommand{\w}{\omega}

\begin{document}

\title{
Fractional impurity moments in two-dimensional non-collinear magnets
}

\author{Alexander Wollny}
\author{Lars Fritz}
\affiliation{
Institut f\"ur Theoretische Physik, Universit\"at zu K\"oln,
Z\"ulpicher Stra\ss e 77, 50937 K\"oln, Germany
}
\author{Matthias Vojta}
\affiliation{Institut f\"ur Theoretische Physik, Technische Universit\"at Dresden,
01062 Dresden, Germany}

\date{\today}

\begin{abstract}
We study dilute magnetic impurities and vacancies in two-dimensional frustrated magnets
with non-collinear order. Taking the triangular-lattice Heisenberg
model as an example, we use quasiclassical methods to determine the impurity
contributions to the magnetization and susceptibility. Most importantly, each impurity
moment is {\em not} quantized, but receives non-universal screening corrections due to
local relief of frustration. At finite temperatures, where bulk long-range order is
absent, this implies an impurity-induced magnetic response of Curie form, with a
prefactor corresponding to a {\em fractional} moment per impurity. We also discuss the
behavior in an applied magnetic field, where we find a singular linear-response limit for
overcompensated impurities, and propose experiments to test our theory.
\end{abstract}

\maketitle

%%%%%%%%%%%%%%%%%%%%%%%%%%%%%%%%%%%%%%%%%%%%%%%%%%%%%%%%%%%%%%%%%%%%%%%

Impurities have been established as a powerful means to both probe and tune bulk properties of
correlated-electron materials.
%
%Prominent examples include the screening of magnetic
%moments in metals (i.e. the Kondo effect), the impurity-induced suppression of the
%critical temperature in superconductors, and the scattering interference of quasiparticles as
%observed in scanning-tunneling spectroscopy.
%
In quantum magnets, non-trivial phenomena include vacancy-induced magnetism in quantum paramagnets
\cite{laflo} and quantum percolation \cite{perco}. Single-impurity behavior has
been predicted to be exotic in quantum critical magnets, where a universal fractional Curie
moment appears at low temperatures \cite{sbv99,sv03,sandvik07}. Isolated impurities in magnets
with long-range order have been studied as well, with most works focussing on the
square-lattice Heisenberg magnet \cite{sbv99,sushkov,sv03,sandvik03}.
%Here, the magnetic properties of a single vacancy have been studied extensively \cite{sbv99,sv03,sandvik03}.

This paper is devoted to impurities in geometrically frustrated spin-$S$ magnets which
order non-collinearly -- a topic which has received little attention
\cite{henley_note}. As we show below, vacancies (i.e. non-magnetic impurities) in
non-collinear magnets display a behavior which is richer and qualitatively different
compared to their collinear counterparts. In particular, the magnetic moment $m$
associated with a single vacancy is {\em not} quantized, in contrast to the collinear
case \cite{sbv99} where it is locked to $m=S$. This effect is already present at the
classical level: nearby spins re-adjust their directions in response to the vacancy,
reflecting that frustration is locally reduced. This partially screens the vacancy
moment, with the screening cloud decaying algebraically due to Goldstone modes.

At zero temperature, the direction of the vacancy moment $m$ is fixed by the bulk
magnetic order. In contrast, at $T>0$ in two dimensions (2d) there is no long-range order
due to the Mermin-Wagner theorem, and the vacancy moment is free to rotate. This rotation
is classical, as it is coupled to a rotation of the bulk spins surrounding the vacancy
\cite{sbv99,sv03}. As a result, the linear-response susceptibility has a singular piece,
$\chi_{\rm imp}(T) = m^2/(3kT)$, corresponding to the Curie response of a {\em
fractional} moment for each vacancy \cite{cut_foot}.
For the triangular-lattice Heisenberg antiferromagnet (AF) with nearest-neighbor interactions,
we find in a $1/S$ expansion
\begin{equation}
m = -0.040 S + 0.196 + \mathcal{O}(1/S) \;,
\label{mres}
\end{equation}
where a negative sign corresponds to overcompensation, described in detail below.
In stark contrast to fractional effective moments found at bulk or boundary
quantum critical points \cite{sbv99,sandvik07,mvrev},
the present mechanism is realized deep inside the renormalized
classical regime \cite{CHN} of a 2d magnet.

A finite magnetic field $h$ has two effects which tend to compete: it orients the impurity
moment parallel to the field and it induces a macroscopic bulk moment. This
bulk--boundary competition is governed by a field-induced length scale $l_h \propto 1/h$
and limits the linear-response regime \cite{eggert07}. For a single overcompensated impurity in
the triangular lattice, we find this competition to be particularly drastic: Linear
response breaks down at any finite field.

In the body of the paper, we sketch the derivation of these results and propose
tests and extensions of the non-trivial screening advocated here. Our considerations
qualitatively apply to a large class of frustrated AFs with non-collinear ground states,
which are unique up to global spin rotations.
For definiteness, we will present results for the spin-$S$ triangular-lattice Heisenberg model
\begin{equation}
\label{H}
\mathcal{H} = \sum_{\langle ij\rangle} \left[
J \vec{S}_i \cdot \vec{S}_j + K (\vec{S}_i \cdot \vec{S}_j)^2
\right] - h \sum_i S_i^z.
\end{equation}
The biquadratic exchange, with its strength parameterized by $k=K/(J S^2)$, generates a
family of models and, in particular, lifts the accidental classical degeneracy of the
simple nearest-neighbor Heisenberg model in an applied field~\cite{Laeuchli}. In zero
field, the ground state is given by the familiar coplanar 120$^\circ$ ordering at
wavevector $\vec Q = (4\pi/3,0)$ for $-2/9<k<2/9$.

%%%%%%%%%%%%%%%%%%%%%%%%%%%%%%%%%%%%%%%%%%%%%%%%%%%%%%%%%%%%%%%%%%%%%%%

{\it Vacancy in the ground state of a classical non-collinear magnet.}
%
%The single-vacancy problem already contains non-trivial physics at the classical level.
Consider a bulk AF with geometric frustration, where not all energetic
constraints (e.g. all neighboring spins pairwise antiparallel) can be satisfied. Removing a
single spin locally reduces frustration due to the elimination of constraints. For a
non-collinear magnet, this seeds a re-adjustment
of spin directions.

\begin{figure}[!t] % FIGURE 1
\includegraphics[width=0.15\textwidth]{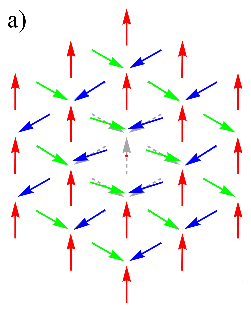}
\hspace*{0.4cm}
\includegraphics[width=0.27\textwidth]{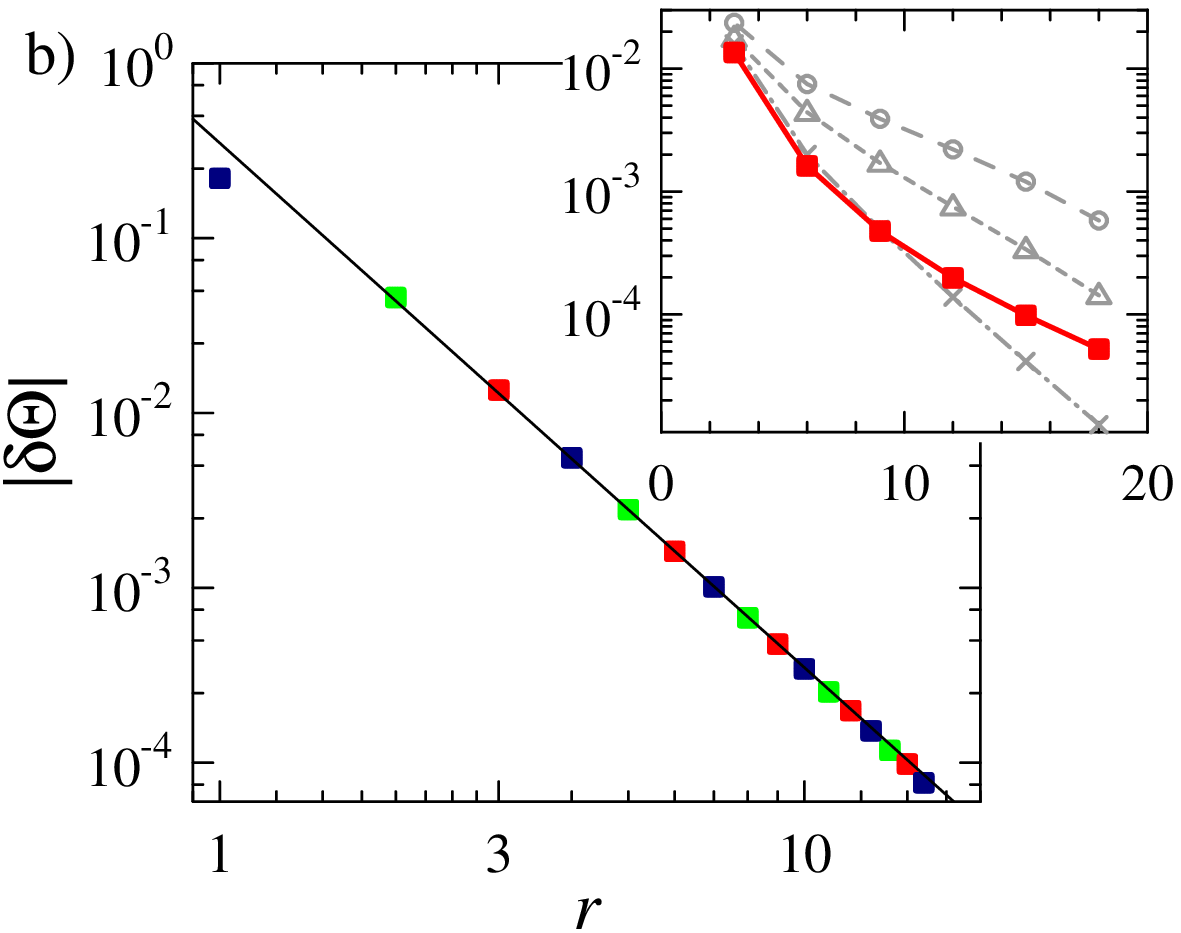}
\caption{
a) Classical ground-state spin configuration of the triangular AF with vacancy,
showing the spin re-adjustment near the vacancy; the dashed arrows indicate the original
$120^\circ$ order.
b) Rotation angle $|\delta \Theta|(r)$ along a high-symmetry line for $k=0$ and $L=51$,
together with the asymptotic power law $\delta \Theta \propto 1/r^3$. The inset shows the
same data (solid), together with $|\delta \Theta(r)|$ at finite field $h/J=0.5$ (dashed),
1.0 (short dash), 1.5 (dash-dot), all for $k=-0.05$, for one sublattice in a log-linear plot. The
exponential (instead of power-law) decay for $h>0$ is obvious.
}
\label{fig:angle}
\end{figure}

For the triangular lattice, this re-adjustment is illustrated in Fig.~\ref{fig:angle}a,
which shows the result of a numerical optimization of the spin directions in a
finite system of size $L^2$ with $L=51$, where a single spin has been removed.
% from the $A$ sublattice.
The spins remain coplanar and rotate by angles $\delta\Theta$ relative to the original
120$^\circ$ configuration, such that the spins near the vacancy tend to be more
antiparallel. The numerical results for $\delta \Theta$ show a sixfold ($f$-wave) angular
symmetry and are consistent with a spatial decay of $\delta \Theta(r) \propto 1/r^3$,
Fig.~\ref{fig:angle}b \cite{henley_note}.

This result is rationalized as follows: The full rotation pattern can be understood
as the response of the system to a field $\tilde h$ which couples to the six neighbors of the
vacancy such that these spins are rotated towards an antiparallel configuration. Hence,
the field $\tilde h$ acting on these six sites is locally transverse and alternating,
in a rotated frame compactly written as
$\tilde h \sum_{j=1}^6 \beta_j S_j^x$ with $\beta_j = (-1)^j$.
The long-distance rotation is determined by a transverse susceptibility, which is
dominated by the modes near the ordering wavevector with linear dispersion $\w_q$.
A straightforward calculation gives
$\delta \Theta(r) \propto \int d^d q e^{i\vec{q}\cdot\vec{r}} \beta_q/\w_q^2 \propto 1/r^{d+1}$.

\begin{figure}[!t]
\includegraphics[width=3.4in]{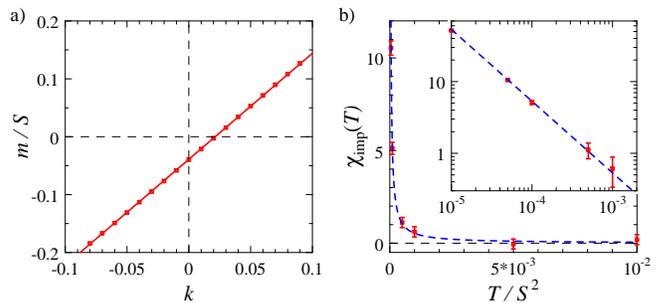}
\caption{
a) Effective vacancy moment $m/S$ for the classical triangular Heisenberg AF as
function of the biquadratic exchange $k$ in Eq.~\eqref{H}.
b) MC results for $\chi_{\rm imp}$ as function of $T/S^2$, calculated with $JS^2=1$ and
$k=0$. The dashed line shows the predicted Curie law $m^2/(3kT)$ \eqref{chiimp} with
$|m/S|=0.04$. The inset shows the low-temperature data in a log-log plot. The data in a (b) have
been obtained for systems of size $L=51$ ($L=9,12$); finite-size effects are negligible.
}
\label{fig:chi}
\end{figure}

The state with a single vacancy has a finite magnetization $m$. While this would simply
be $m=S$ without re-adjusted angles (i.e. in the collinear case), the re-adjustment tends
to screen this moment. For the triangular lattice, the numerical result, obtained from
integration over the screening cloud, is $m/S=-0.0396(3)$, i.e., the missing spin is {\em
over}compensated, such that the total moment points in the direction of the removed spin.
The value of $m$ is non-universal, i.e., depends on details of the Hamiltonian:
Fig.~\ref{fig:chi}a shows $m/S$ as function of the biquadratic exchange coupling
$K$ in Eq.~\eqref{H}.

%%%%%%%%%%%%%%%%%%%%%%%%%%%%%%%%%%%%%%%%%%%%%%%%%%%%%%%%%%%%%%%%%%%%%%%

{\it Vacancy: $1/S$ corrections.}
Quantum corrections to the classical $T=0$ results can be obtained using
spin-wave theory. Holstein-Primakoff bosons $a$ are introduced to capture deviations
from the classical state in the presence of a vacancy, Fig.~\ref{fig:angle}a.
Upon expressing the Heisenberg model in terms of the $a$ bosons, terms linear in $a$
vanish as required. Linear
spin-wave theory amounts to a diagonalization of the quadratic-in-$a$ piece of the
Hamiltonian, which has to be done numerically for finite lattices \cite{wessel}
due to the inhomogeneous reference state. Doing so, we find the
full spectrum of eigenvalues and eigenvectors, which can be used to calculate
$1/S$ corrections to thermodynamic observables as well as response functions.
% As usual, the spin directions $\Theta(\vec{r}_i)$ are not changed to order $1/S$.

The local magnetization correction $\delta m(\vec{r}_i) = \langle a_i^\dagger a_i
\rangle$ decays to the known bulk value of $\delta m_b = 0.26$ \cite{cherny} at long
distances, corresponding to a staggered magnetization of $m_b = S - 0.26$.
The impurity contribution, $\delta m(\vec{r}_i) - \delta m_b$, indicates enhanced quantum corrections near the impurity which fall off as $1/r^3$, consistent with the
Goldstone-mode expectation. The $1/S$ correction to the uniform moment associated with
the vacancy is obtained from integration, $\delta m = \sum_i \delta m(\vec{r}_i) \cos
\Theta(\vec{r}_i)$, which evaluates to $\delta m = 0.196$, Eq.~\eqref{mres}. Further
corrections at higher orders in $1/S$ will not qualitatively modify the result of a
non-universal fractional value of $m$, but apparently both overcompensation and
undercompensation may occur, depending on $S$ and microscopic details.
We note that local impurity-induced magnetization corrections obeying $\delta
m(\vec{r}_i)-\delta m_b \propto 1/r^3$ also occur in the collinear square-lattice case, but here spin
conservation demands that the integral $\delta m$ vanishes; hence $m$ remains locked to
$S$ \cite{sushkov,sv03}.

%Careful with $\chi$ and $S\to\infty$.
%
%In general, finite bulk $\chi$ is $\mathcal{O}(S^0)$.
%
%But free spin $\chi$ is $S(S+1)/(3k_B T)$!
%Scaling with $S$ works because $T\sim E\sim S^2$.
%
%Clarify: Which order in $1/S$ gives non-zero tranverse/longitudinal response
%with/without vacancy? In spin waves, chi-long-collinear is bubble whereas
%chi-perp-collinear is single-pt propagator.

%%%%%%%%%%%%%%%%%%%%%%%%%%%%%%%%%%%%%%%%%%%%%%%%%%%%%%%%%%%%%%%%%%%%%%%

{\it Finite temperatures: Fractional Curie response.}
For $T>0$ in two space dimensions, % and in the absence of spin anisotropies,
long-range bulk magnetic order is destroyed by thermal fluctuations, with the correlation
length $\xi$ being exponentially large at low temperatures, $T\ll J$.
Consequently, the direction of the impurity moment is no longer fixed, but is free to rotate with
the local orientation of the bulk magnetic domain surrounding the impurity. It has been
shown both analytically \cite{sbv99,sv03} and numerically \cite{sandvik03} that this
rotation is classical and leads to a linear response of Curie form:
\begin{equation}
\label{chiimp}
\chi_{\rm imp}(T) = \frac{m^2}{3kT} + \mathcal{O}(T^0) \;,
\end{equation}
where the subleading term receives a multiplicative logarithmic correction in 2d
\cite{sandvik03,sv03,nagaosa}.

In the non-collinear case, the partial screening of the vacancy moment, established above
for $T=0$, will remain intact at small $T>0$ because of the large correlation length.
This implies a Curie response \eqref{chiimp} corresponding to a fractional moment per vacancy.
This central result is fully borne out by numerics: we have performed classical Monte
Carlo simulations of triangular-lattice Heisenberg magnets, using the standard Metropolis
algorithm. In Fig.~\ref{fig:chi}b we show the result for the impurity susceptibility
$\chi_{\rm imp}$, obtained from subtracting the linear-response $\chi$ of a system with
vacancy from that of a system without vacancy. While the high-temperature part is
difficult to analyze given the error bars, the low-temperature data clearly show a Curie
divergence, with a prefactor consistent with $|m/S|=0.04$ within error bars.
For $S<\infty$, we expect a subleading $\log T$ contribution to $\chi_{\rm imp}(T)$
arising from Goldstone modes similar to the collinear case; a detailed analysis will
appear elsewhere.

%%%%%%%%%%%%%%%%%%%%%%%%%%%%%%%%%%%%%%%%%%%%%%%%%%%%%%%%%%%%%%%%%%%%%%%

{\it Vacancy vs. extra spin.}
So far, we have considered the special case of a vacancy, experimentally obtained by
replacing a magnetic by a non-magnetic ion. A different type of impurity is an extra spin
of size $S'$, coupled to a single site of the bulk magnet with a Heisenberg coupling $J'$.
For antiferromagnetic $J'\gg J$ and $S=S'$, the impurity spin and its bulk partner lock
into a singlet, and we recover the vacancy case. On the other hand, for $J'\ll J$ the
re-adjustment of the spin directions due to the impurity will be parametrically small in
$J'/J$, and we expect for the impurity moment $m\to S$ as $J'\to 0$.
Hence, varying $J'/J$ leads to a continuous change of $m$.
Interestingly, this crossover {\em cannot} be captured in a $1/S$ expansion for the
extra-spin problem: In the classical limit, there is no singlet formation for large $J'$,
and this is not recovered at any order in $1/S$, as can be seen by explicit calculation.
%In other words, the limits $J'/J\to\infty$ and $S\to\infty$ do not commute.
%In fact, at
%order $1/S$ there is no re-adjustment of spin directions at any finite $J'$, and a $1/S$
%analysis is only useful in the small-$J'$ limit.
However, the direct vacancy calculation {\em can} be performed in $1/S$, as shown above,
the crucial difference being that the missing spin (or the $J'=\infty$ singlet) is built
in from the outset.

%%%%%%%%%%%%%%%%%%%%%%%%%%%%%%%%%%%%%%%%%%%%%%%%%%%%%%%%%%%%%%%%%%%%%%%

\begin{figure}
\includegraphics[width=0.28\textwidth]{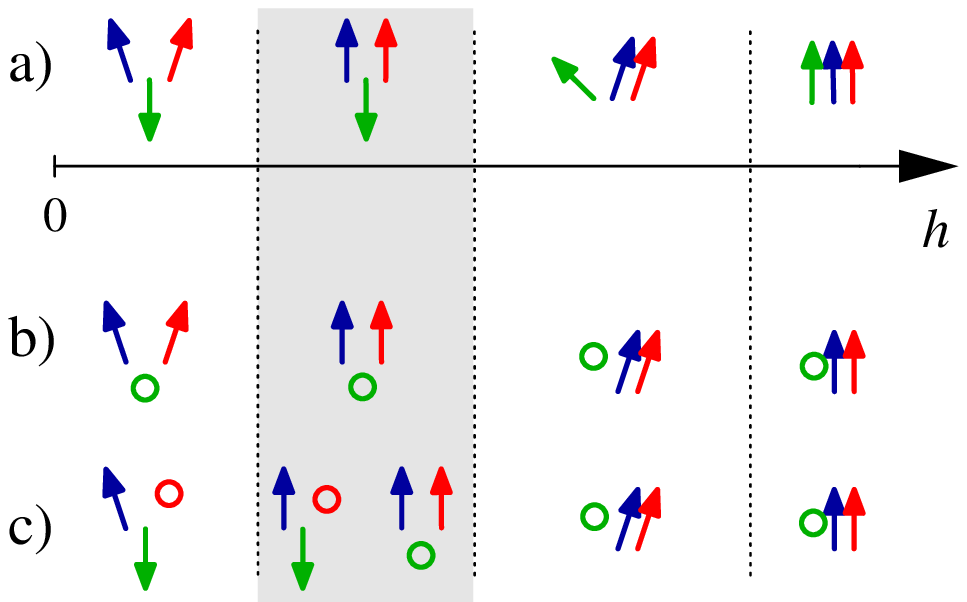}
\hspace*{0.5cm}
\includegraphics[width=0.15\textwidth]{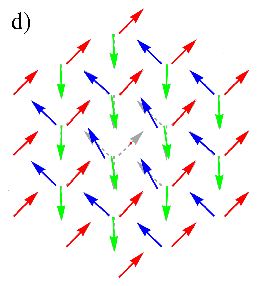}
\caption{
a) Schematic evolution of the three-sublattice coplanar bulk spin configurations as
function of applied field for the triangular AF, with the 1/3 magnetization plateau shaded.
These states are selected out of the classical $k\!=\!0$ ground-state manifold of
$\mathcal{H}$ \eqref{H} by both thermal and quantum fluctuations as well as
negative small $k$.
A single vacancy chooses one of the sublattices: b) undercompensated and c)
overcompensated case. In b), an impurity quantum phase transition occurs inside the
plateau phase. (Note that undercompensation does not occur in our family of classical
models with $k<0$, but is expected for $S<\infty$ from Eq.~\eqref{mres}.)
d) Spin configuration with (overcompensated) vacancy, calculated for $h/J=1.0$, $k=-0.05$, and $L=51$.
}
\label{fig:hschem}
\end{figure}

{\it Finite magnetic field.}
For the collinear square-lattice AF, it has been shown that a vacancy in a finite applied
field generates spin textures in its vicinity, which result from the competition between
aligning the vacancy moment and inducing a bulk moment \cite{eggert07}.
Here we investigate the non-collinear case on the triangular lattice. As the
nearest-neighbor Heisenberg model has an accidental degeneracy of classical
ground states at finite fields, which is lifted in favor of coplanar states
(Fig.~\ref{fig:hschem}a) both by quantum and thermal fluctuations \cite{chubukov,kawa85}, we
choose to investigate the classical model with biquadratic exchange, Eq.~\eqref{H} with
$-2/9<k<0$, which leads to the same coplanar finite-field phases as the ones selected by
fluctuation effects. (Non-coplanar states are favored for $0<k<2/9$.)

Qualitatively, the vacancy physics strongly differs between the undercompensated and
overcompensated cases. For undercompensation, Fig.~\ref{fig:hschem}b, a small field will
orient the system such that the vacancy sublattice points antiparallel to the field. This
is compatible with the field-induced bulk state, hence a strong competition between bulk
and boundary effects is absent, and the zero-field limit will be smooth.

\begin{figure}[!t]
\includegraphics[width=3.4in]{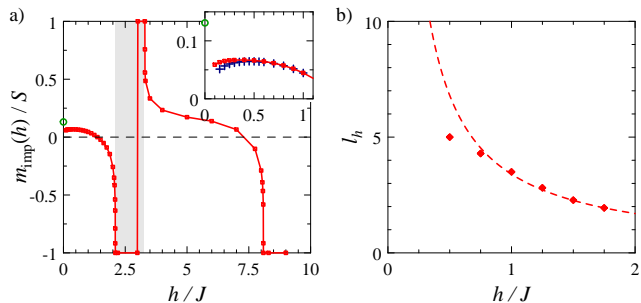}
\caption{
a) Impurity contribution to the magnetization, $m_{\rm imp}(h)/S$, as function of applied field
$h$, for the classical triangular AF \eqref{H} with $k=-0.05$. The shaded region
corresponds to the bulk 1/3 magnetization plateau. The inset shows a zoom onto the
small-field region; squares (crosses) are data for $L=51$ ($L=21$). The circle at $h=0$
represents the linear-response value $|m/S|$, demonstrating the breakdown of linear
response for this overcompensated case.
b) Field-induced length scale $l_h(h)$ obtained from an exponential fit to $\delta\Theta(r)$ for
$L=51$, together with the anticipated $l_h \propto 1/h$ behavior (dashed). Finite-size
effects are important for small $h>0$ where $l_h \ll L$ is violated.
}
\label{fig:field}
\end{figure}

This is different in the overcompensated case, where orienting the vacancy moment in
field direction is {\em incompatible} with the bulk state. Our numerics shows that the
system chooses a compromise such that the vacancy sits in one of the sublattices directed
approximately parallel to the field, with a significant distortion near the vacancy,
Fig.~\ref{fig:hschem}c,d. This distortion falls off exponentially (there is no coupling
to the remaining Goldstone mode), with a length scale $l_h \propto 1/h$ \cite{eggert07},
Fig.~\ref{fig:field}b.
Most importantly, the zero-field limit is {\em singular} in this case, i.e., the
distortion pattern for $h\!\to\!0$ does not recover its zero-field structure. This is seen in
both the inset of Fig.~\ref{fig:angle}b and Fig.~\ref{fig:field}a. The latter shows
%the impurity contribution to the uniform magnetization,
$m_{\rm imp}(h)$, defined as the difference of the total magnetizations with and without
vacancy. By construction, $m_{\rm imp}(h\!=\!0) = |m|$ and $m_{\rm imp}(h\!\to\!\infty)=-S$.
Fig.~\ref{fig:field}a demonstrates that $m_{\rm imp}(h\!\to\!0)$ again represents a fractional
impurity moment which is different from $|m|$ in the overcompensated case.

The evolution of $m_{\rm imp}(h)$ through the bulk magnetization plateau is also very
different in the undercompensated and overcompensated cases, Fig.~\ref{fig:hschem}b,c.
While it is smooth for undercompensation, a jump occurs at $h=3J$ in the overcompensated
case, Fig.~\ref{fig:field}a. This signals a first-order impurity transition where the vacancy site
switches the sublattice implying that the entire bulk configuration rearranges.

%%%%%%%%%%%%%%%%%%%%%%%%%%%%%%%%%%%%%%%%%%%%%%%%%%%%%%%%%%%%%%%%%%%%%%%

{\em Finite impurity concentration.}
We finally discuss the measurable consequences of our findings in the realistic case of a
finite impurity concentration $n_{\rm imp}$. Assuming that the impurities are distributed
equally over all sublattices, their moments tend to average out at $T=0$ and $h=0$. This
behavior persists at finite $T$, provided that $\xi \gg l_{\rm imp}$ where $l_{\rm imp} =
n_{\rm imp}^{1/d}$ is the mean impurity distance. In the opposite limit, $\xi \ll l_{\rm
imp}$, the impurity moments fluctuate independently, and their response simply adds
up. Hence, observing fractional Curie response of independent impurity moments is
possible at elevated $T$ and small $n_{\rm imp}$ \cite{cut_foot}. Note that elevated
fields which induce $l_h \ll l_{\rm imp}$ also lead to an effective decoupling of
multiple impurity moments, which, however, are polarized in this limit.
The spin re-arrangement predicted to occur inside the plateau phase for overcompensated
impurities is detectable by local probes like NMR.

%%%%%%%%%%%%%%%%%%%%%%%%%%%%%%%%%%%%%%%%%%%%%%%%%%%%%%%%%%%%%%%%%%%%%%%

{\it Conclusions.}
For impurities in non-collinear magnets, our main result is a partial screening of the
impurity magnetic moment, leading to a fractional Curie response at low temperatures in
the 2d case. We have evaluated the vacancy moment for the spin-$S$ triangular-lattice AF
in a $1/S$ expansion, but we expect our qualitative results to be valid for any
frustrated AF with non-collinear ground state (which is unique up to global spin
rotations).

Our predictions could in principle be verified by large-scale numerical studies in
analogy to Refs.~\cite{sandvik03,sandvik07}, however, quantum Monte-Carlo
approaches are plagued by the sign problem which is serious for most frustrated
AF.

On the experimental side, one can expect the physics described here to be
generically realized, as Curie tails in $\chi(T)$ due to impurities are routinely
observed in magnets. A {\em quantitative} analysis of these tails in samples with known
concentration of impurities would allow to extract the fractional moment size $m$ (in a
regime where interactions between the impurity moments are small); our prediction is
$m \ll S$ in contrast to the behavior in collinear magnets.

An interesting open question is how the fractional moment advocated here evolves upon
approaching a quantum critical point of the bulk magnet, where {\em at} criticality a
universal fractional response is expected.
Our results also call for investigations of vacancies in frustrated collinear magnets
where vacancies may induce non-collinear spin textures in order to reduce frustration.

%%%%%%%%%%%%%%%%%%%%%%%%%%%%%%%%%%%%%%%%%%%%%%%%%%%%%%%%%%%%%%%%%%%%%%%

% \acknowledgments

We thank R. Moessner, A. Rosch, Q. Si, and O. A. Starykh for illuminating discussions.
This research was supported by the DFG through SFB 608 (K\"oln) and GRK 1621 (Dresden).
MV also acknowledges financial support by the Heinrich-Hertz-Stiftung NRW and the
hospitality of the Centro Atomico Bariloche. % where part of this work was performed.

%%%%%%%%%%%%%%%%%%%%%%%%%%%%%%%%%%%%%%%%%%%%%%%%%%%%%%%%%%%%%%%%%%%%%%%


\begin{thebibliography}{}
\bibitem{cut_foot}
Practically, the Curie response will be cut-off by interactions between the vacancy moments
and by spin anisotropies leading to bulk order at finite $T$.
\end{thebibliography}
\end{document}